\documentclass[onecolumn,aps,prb,showpacs,amsfonts,amssymb,amsmath]{revtex4}
\usepackage{mathrsfs}
\usepackage{amsmath}
\usepackage{graphicx}

\begin{document}
 %\draft
 \title{Cooperative order and excitation spectra in the bicomponent spin networks}
 \author{Bao Xu, Han-Ting Wang, and Yupeng Wang}
 \address{Beijing National Laboratory for Condensed Matter
 Physics, Institute of Physics, Chinese Academy of Sciences, Beijing
 100080, China}

\begin{abstract}
 A ferrimagnetic spin model composed of $S=\frac{1}{2}$ spin-dimers and $S=\frac{5}{2}$
 spin-chains is studied by combining
 the bond-operator representation (for $S=\frac{1}{2}$ spin-dimers) and
 Holstein-Primakoff transformation (for $S=\frac{5}{2}$ spins).
 A finite interaction $J_{\rm DF}$ between the spin-dimer and the spin chain makes the spin chains ordered
 antiferromagnetically and the spin dimers polarized. The effective interaction between the spin chains,
 mediated by the spin dimers, is calculated up to the third order. The staggered magnetization in the spin dimer is
 shown proportional to $J_{\rm DF}$. It presents an effective staggered field reacting on the spin chains.
 The degeneracy of the triplons is lifted due to the chain magnetization and a mode with longitudinal polarization
 is identified. Due to the triplon-magnon interaction, the hybridized triplon-like excitations show different
 behaviors near the vanishing $J_{\rm DF}$. On the other hand, the hybridized magnon-like excitations open a gap
 $\Delta_A\sim J_{\rm DF}$. These results consist well with the experiments on Cu$_{2}$Fe$_{2}$Ge$_{4}$O$_{13}$.
\end{abstract}


%\submitto{\EJP}

 \maketitle

\section{Introduction}

 Quantum magnetism has received considerable attention from both
 theoretical and experimental points of view in the past decades.
 Some low-dimensional magnets, for example,
 antiferromagnetic spin chains with half odd integer spins are gapless and have
 a disordered ground state; while some others, such as antiferromagnetic
 spin chains with integer spins, spin ladders and dimerized spin chains, are
 gapped and disordered. In two dimensional cases, Heisenberg antiferromagnets in the square
 lattice are gapless and have ordered ground states at zero temperature.
 Besides the dimensional effect, various frustration and anisotropy cause novel and complex phenomena.
 Recently, bicomponent systems combining two different spin frameworks have been realized experimentally
 in R$_2$BaNiO$_5$\cite{zheludev96,zheludev98},
 Cu$_2$Fe$_2$Ge$_4$O$_{13}$\cite{masuda2003,masuda2004,masuda2005,masuda2007},
 Cu$_2$CdB$_2$O$_6$\cite{hase2005} and
 Cu$_3$Mo$_2$O$_9$\cite{hamasaki2008}.
 Among them, Cu$_{2}$Fe$_{2}$Ge$_{4}$O$_{13}$, which incorporates intercalated
 Cu$^{2+}$ spin dimers ($S={1\over 2}$) and Fe$^{3+}$ spin chains ($S={5\over 2}$),
 is most extensively studied. Below $T_N=39$ K, a cooperative order was observed by the measurements of susceptibility
 and heat capacity\cite{masuda2004}. At $T=1.5$ K, the estimated $m_{\rm Cu}=0.38$ $\mu_B$ and $m_{\rm Fe}=3.62$ $\mu_B$.
 Compared to the classical expectation value of $1$ $\mu_B$, $m_{\rm Cu}$ is drastically suppressed and keeps
 proportional to $m_{\rm Fe}$ at all temperatures.
 By detailed inelastic neutron scattering study\cite{masuda2004,masuda2005,masuda2007},
 Cu$_{2}$Fe$_{2}$Ge$_{4}$O$_{13}$ was found to exhibit two types of spin excitations with separate energy scales.
 Although the Fe$^{3+}$-centered low-energy spin excitations can be well interpreted by the spin wave theory
 and a small gap of about $1$ meV is estimated, the Cu$^{2+}$-centered high-energy part is less understood.
 Masuda et. al\cite{masuda2004} guessed the presence of a triplet mode, which should have longitudinal
 polarization and be totally incompatible with conventional spin wave theory. In this paper, we combine
 bond operator representation\cite{sachdev} and Holstein-Primakoff transformation\cite{holstein1940}
 to study this bicomponent system.

\begin{figure}
 \includegraphics[ scale=0.70 ]{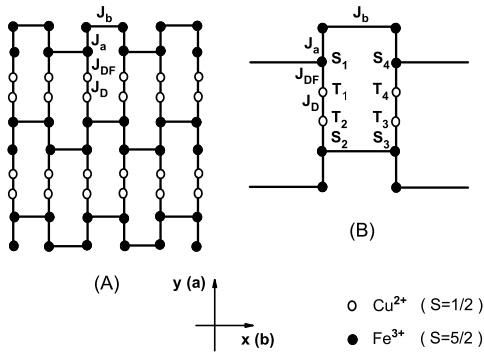}\\
 \caption{The geometry of the topologically equivalent spin lattice of Cu$_{2}$Fe$_{2}$Ge$_{4}$O$_{13}$.
  The Cu-Cu interaction $J_{\rm D}$ is dominantly larger than the Fe-Fe interaction $J_{\rm a(b)}$ and the
  Cu-Fe interaction $J_{\rm DF}$. The weak interchain interaction along c-direction
  is neglected for simplicity.}\label{str2}
 \end{figure}

Since the exchange coupling along the c-axis is pretty weak, the real geometry of
 Cu$_{2}$Fe$_{2}$Ge$_{4}$O$_{13}$ could be simplified as a
 two-dimensional topologically equivalent model\cite{masuda2003}
 (Fig.\ref{str2}). The model Hamiltonian consists of the Cu-Cu, the Fe-Fe and the Cu-Fe
 interactions:
 \begin{eqnarray}
 H=H_{\rm Cu}+H_{\rm Fe}+H_{\rm Cu-Fe},
 \end{eqnarray}
 with
 \begin{eqnarray}
 H_{\rm Cu}&=&\sum_{\vec r} J_{\rm D}(\hat{T}_{\vec r,1}\cdot\hat{T}_{\vec r,2}
    +\hat{T}_{\vec r,3}\cdot\hat{T}_{\vec r,4}),\nonumber\\
 H_{\rm Fe}&=&\sum_{\vec r} \frac{1}{2}J_{\rm a}(\hat{S}_{\vec r,1}
    \cdot\hat{S}_{\vec r+\vec a,2}+\hat{S}_{\vec r +\vec a,3}
    \cdot\hat{S}_{\vec r,4}\nonumber\\
 &&+\hat{S}_{\vec r -\vec a,1}
    \cdot\hat{S}_{\vec r,2}+\hat{S}_{\vec r,3}
    \cdot\hat{S}_{\vec r -\vec a,4}) \nonumber\\
 &&+J_{\rm b}(\hat{S}_{\vec r,3}
    \cdot\hat{S}_{\vec r,2}+\frac{1}{2}\hat{S}_{\vec r,1}
    \cdot\hat{S}_{\vec r-\vec b,4}+\frac{1}{2}\hat{S}_{\vec r+\vec b,1}
    \cdot\hat{S}_{\vec r,4}),\nonumber\\
  H_{\rm Cu-Fe}&=&\sum_{\vec r,i=1,2,3,4} J_{\rm DF}\hat{T}_{\vec r,i}\cdot\hat{S}_{\vec r,i},\label{model}
 \end{eqnarray}
 where $J_{\rm D}$ and $J_{\rm DF}$ denote the
 Cu-Cu and Cu-Fe interactions respectively, and $J_{\rm a(b)}$ the Fe-Fe exchange constant along a(b) direction.
 By neutron inelastic scattering and neutron diffraction\cite{masuda2007}, $J_{\rm D}$ was found dominantly larger
 than other interactions and $J_{\rm b}$ is very close to $J_{\rm a}$.
 The Fe$^{3+}$ has spin $S=\frac{5}{2}$ and the Cu$^{2+}$ has spin $T=\frac{1}{2}$.
 Single-ion anisotropy and external magnetic field are not considered here. With a vanishing $J_{\rm DF}$, the model
 reduces to the independent one-dimensional Heisenberg antiferromagnetic chains with $S=\frac{5}{2}$ and isolated spin
 dimers with $T=\frac{1}{2}$. With a finite $J_{\rm DF}$, an effective interaction between the Fe$^{3+}$ chains, mediated
 by the Cu$^{2+}$ dimers, will make the Fe$^{3+}$ spins stay in the N\'{e}el state.
 (The small Fe-Fe interactions along c-direction reinforce this order). On the other hand, the local magnetization
 of Fe$^{3+}$ spins will tend to excite the Cu$^{2+}$ dimers from the singlet to the triplet
 and the Cu$^{2+}$ spins will then show local magnetizations.
 In the following, we study the effective interaction between the Fe$^{3+}$ chains mediated by the Cu$^{2+}$ dimers
 perturbatively and the Cu$^{2+}$ magnetization induced by the local Fe$^{3+}$ spins(section II). In section III, we study
 the excitation spectra of the mixed spin system. A summary is given in section IV.

 \section{Indirect Fe-Fe magnetic interaction and Cu$^{2+}$ magnetization polarized by the Fe$^{3+}$ spin}

As mentioned before, $J_{\rm D}$ is much larger than $J_{\rm a(b)}$
 and $J_{\rm DF}$. Considering the low energy
 excitations only and integrating out the degrees of freedom
 of the spin dimers, we could obtain an effective Fe-Fe exchange
 $J_{\rm eff}$ mediated by isolated spin-$\frac{1}{2}$ dimers\cite{wang2001}.
 In the model Hamiltonian(eq. (1)), $H_{\rm Cu-Fe}$ could be treated as a small amount and $H_{\rm Fe}$ will be neglected
 for the moment since no Cu$^{2+}$ spin operators are involved in it. As shown in Fig. 2,
 the model then reduces to a four-spin system
 \begin{eqnarray}
 {\cal H}=H_0+H^{\prime}
 \end{eqnarray}
 with $H_0=J_{\rm D}\hat{T}_1\cdot \hat{T}_2$
 and $H^{\prime}=J_{\rm DF}(\hat{T}_1\cdot\hat{S}_1+\hat{T}_2\cdot\hat{S}_2)$.

 \begin{figure}
 \includegraphics[ scale=0.70 ]{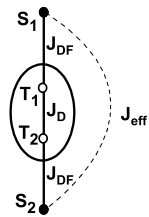}\\
 \caption{The effective four-spin model.}\label{pertmodel}
 \end{figure}

 $H_0$ has a singlet ground state $|s>$ with a ground state energy
 $E_{0}=-\frac{3}{4}J_{\rm D}$.
 Its excited states are triplets $|t_{\alpha}>$ with $\alpha=0,\pm 1$, corresponding to the total
 $z$ component $T_{1z}+T_{2z}=0,\pm 1$. The eigenenergy of the excited states is
 $E_{1,\alpha}=\frac{1}{4}J_{\rm D}$. The first order perturbation of $H_{\rm Cu-Fe}$ is
 ${\cal H}_1=<s|H_{\rm Cu-Fe}|s>=0$. The 2nd-order perturbation
 ${\cal H}_2=\sum_{\alpha} {<s|H_{\rm Cu-Fe}|t_{\alpha}><t_{\alpha}|H_{\rm Cu-Fe}|s>\over {E_0-E_{1\alpha}}}
 ={J_{\rm DF}^2\over 2J_{\rm D}} \hat{S}_{1}\cdot\hat{S}_{2}$. The 3rd-order perturbation is
 ${\cal H}_3=\sum_{\alpha,\beta}{<s|H_{\rm Cu-Fe}|t_{\alpha}><t_{\alpha}|H_{\rm Cu-Fe}|t_{\beta}><t_{\beta}|H_{\rm Cu-Fe}|s>
 \over (E_0-E_{1,\alpha})(E_0-E_{1,\beta})}={3J_{\rm DF}^3\over 4J_{\rm D}^2} \hat{S}_{1}\cdot\hat{S}_{2}$.
 Thus, up to the third order, the effective exchange coupling mediated by the spin-$1/2$ dimer is obtained as
\begin{eqnarray}
 {\cal H}_{\rm eff}&=&J_{\rm eff}\hat{S}_{1}\cdot\hat{S}_{2}
 \end{eqnarray}
 with
 $J_{\rm eff}=\frac{J^{2}_{\rm DF}}{2J_{\rm D}}\left(1+\frac{3J_{\rm DF}}{2J_{\rm D}}\right)$.

With the effective interaction between the Fe$^{3+}$ spin chains considered, the Fe$^{3+}$ spins will
 order antiferromagnetically. We use a molecular field approximation to study the effects of
 the staggered Fe$^{3+}$ magnetization on the Cu$^{2+}$ spin dimers. It is relevant at a higher energy
 range (comparable to the dimer gap), where the dynamics of the system is dominated by the dimers and the
 chain freedom can be effectively integrated out.
 The four spin Hamiltonian becomes
 \begin{eqnarray}
 {\cal H}=J_{\rm D}\hat{T}_{1}\cdot\hat{T}_{2}+J_{\rm DF}m_{\rm Fe}({T}_{1z}-{T}_{2z}),
 \end{eqnarray}
 where $m_{\rm Fe}=<S_z>$ is the staggered magnetization of Fe$^{3+}$ spin. This approximate Hamiltonian can be
 exactly diagonalized with the bases $|s>$ and $|t_{\alpha}>$ ($\alpha=0,\pm 1$). The eigenvalues are
 $e_0=-{1\over 4}J_{\rm D}-{1\over 2}J_{\rm D}\sqrt {1+{4J_{\rm DF}^2m_{\rm Fe}^2\over J_{\rm D}^2}}$,
 $e_1=-{1\over 4}J_{\rm D}+{1\over 2}J_{\rm D}\sqrt {1+{4J_{\rm DF}^2m_{\rm Fe}^2\over J_{\rm D}^2}}$,
 and $e_2=e_3={1\over 4}J_{\rm D}$. The corresponding eigenstates are $|\psi_i>=a_i|s>+b_i|t_0>$ ($i=0,1$) with
 $a_i={J_{\rm DF}m_{\rm Fe}\over\sqrt{(e_i+{3\over 4}J_{\rm D})^2+J_{\rm DF}^2m_{\rm Fe}^2}}$ and
 $b_i={e_i+{3\over 4}J_{\rm D}\over\sqrt{(e_i+{3\over 4}J_{\rm D})^2+J_{\rm DF}^2m_{\rm Fe}^2}}$,
 $|\psi_3>=|t_1>$ and $|\psi_4>=|t_{-1}>$. From the ground state $|\psi_0>$,
 we find that $|t_0>$ is partly excited and the Cu$^{2+}$
 spins now get a finite staggered magnetization: $m_{\rm Cu}=<T_{2z}>=-<T_{1z}>=<\psi_0|T_{2z}|\psi_0>=a_0b_0$. When
 $m_{\rm Fe}$ or $J_{\rm DF}\over J_{\rm D}$ is small, we have $m_{\rm Cu}\approx {J_{\rm DF}\over J_{\rm D}}m_{\rm Fe}$, which
 agree well with the experiment\cite{masuda2004}.

\section{Excitations in the bicomponent system}

In the Fe-Cu spin system, if we neglect the interaction between the
 Cu$^{2+}$ and Fe$^{3+}$ spins, the Fe$^{3+}$ spins will be in the N\'{e}el
 state (considering the antiferromagnetic interaction along
 c-direction) and its excitations are spin-waves (magnons); while the
 Cu$^{2+}$ dimers are in the singlet states and the excitations are
 triplets (triplons). When the Cu-Fe interactions are switched on,
 the N\'{e}el order in the Fe$^{3+}$ spin network is reinforced and a
 staggered magnetization in the Cu$^{2+}$ spins are induced.
 The magnon and triplon interact with each other and the hybridized excitations
 show interesting behavior. In the following, we study these excitations.

 For Cu$^{2+}$ spin dimers, we use the bond operator representation.
 With the definition of $|s>=s^{\dag}|0>$, $|t_0>=t^{\dag}|0>$,
 $|t_{-1}>=d^{\dag}|0>$, $|t_1>=u^{\dag}|0>$ and
 $s^{\dag}s+t^{\dag}t+d^{\dag}d+u^{\dag}u=1$, the Cu$^{2+}$ spin
 operators are expressed as\cite{sachdev}:
 \begin{eqnarray}
 T^{+}_{\sigma} &=& \frac{1}{\sqrt{2}}
    [-(u^{\dagger}_{\sigma}s_{\sigma}-s^{\dagger}_{\sigma}d_{\sigma})
    +t^{\dagger}_{\sigma}d_{\sigma}+u^{\dagger}_{\sigma}t_{\sigma}];\nonumber\\
 T^{+}_{\sigma+1} &=& \frac{1}{\sqrt{2}}
    [(u^{\dagger}_{\sigma}s_{\sigma}-s^{\dagger}_{\sigma}d_{\sigma})
    +t^{\dagger}_{\sigma}d_{\sigma}+u^{\dagger}_{\sigma}t_{\sigma}];\nonumber\\
 T^{z}_{\sigma} &=&
    \frac{1}{2}[(s^{\dagger}_{\sigma}t_{\sigma}+t^{\dagger}_{\sigma}s_{\sigma})
    +u^{\dagger}_{\sigma}u_{\sigma}-d^{\dagger}_{\sigma}d_{\sigma}];\nonumber\\
 T^{z}_{\sigma+1} &=&
    \frac{1}{2}[-(s^{\dagger}_{\sigma}t_{\sigma}+t^{\dagger}_{\sigma}s_{\sigma})
    +u^{\dagger}_{\sigma}u_{\sigma}-d^{\dagger}_{\sigma}d_{\sigma}];
 \end{eqnarray}
 where, $\sigma=1,3$, corresponding to the pairs of $({\hat T}_1,{\hat T}_2)$ and $({\hat T}_3,{\hat T}_4)$,
 respectively.

We apply Holstein-Primakoff transformation to Fe$^{3+}$ spins\cite{holstein1940}.
 Supposing  ${\hat S}_1$ and ${\hat S}_3$ point up and ${\hat S}_2$ and ${\hat S}_4$ point down,
 we have
 \begin{eqnarray}
 S^{+}_{i}&=&\sqrt{2S-a^{\dagger}_{i}a_{i}}a_{i},\nonumber\\
 S^{-}_{i}&=&a_i^{\dag}\sqrt{2S-a^{\dagger}_{i}a_{i}},\nonumber\\
 S^{z}_{i}&=&S-a^{\dagger}_{i}a_{i},~~~~i=1,3;
 \end{eqnarray}
 and
 \begin{eqnarray}
 S^{+}_{j}&=&b^{\dagger}_{j}\sqrt{2S-b^{\dagger}_{j}b_{j}},\nonumber\\
 S^{-}_{j}&=&\sqrt{2S-b^{\dagger}_{j}b_{j}}b_{j},\nonumber\\
 S^{z}_{j}&=&b^{\dagger}_{j}b_{j}-S,~~~~j=2,4.
 \end{eqnarray}

Before we substitute these representations into Hamiltonian (1), some approximations have to be made. As found
 in section II, the Cu$^{2+}$ dimers are in a mixed state of $|s>$ and $|t_0>$. We introduce another four
 operators as\cite{vojta1999,vojta2001}
 \begin{eqnarray}
 \left(\begin{array}{c}X_{\sigma}\\Y_{\sigma}\end{array}\right)
    =\left(\begin{array}{cc}{\rm cos}\theta&-{\rm sin}\theta
    \\{\rm sin}\theta&{\rm cos}\theta\end{array}\right)
    \left(\begin{array}{c}s_{\sigma}\\t_{\sigma}\end{array}\right)
 \end{eqnarray}
    and
 \begin{eqnarray}
 \left(\begin{array}{c}s_{\sigma}\\t_{\sigma}\end{array}\right)
    =\left(\begin{array}{cc}{\rm cos}\theta&{\rm sin}\theta
        \\-{\rm sin}\theta&{\rm cos}\theta\end{array}\right)
    \left(\begin{array}{c}X_{\sigma}\\Y_{\sigma}\end{array}\right),
 \end{eqnarray}
 with $\sigma=1,3$ and suppose $X_{\sigma}$ bosons are condensed with $<X_{\sigma}>=<X^{\dag}_{\sigma}>=1$, which
 means a long range order in the Cu$^{2+}$ spin network. $\theta$ will be determined variationally or by canceling
 the single-operator terms.

 For the Holstein-Primakoff transformation, we employ the usual linear approximation: $S^{+}_{i}=\sqrt{2S}a_{i}$,
 $S^{-}_{i}=\sqrt{2S}a_i^{\dag}$, $S^{z}_{i} =S-a^{\dagger}_{i}a_{i}$, $i=1,3$ and $S^{+}_{j}=\sqrt{2S}b_{j}^{\dag}$,
 $S^{-}_{j}=\sqrt{2S}b_j$, $S^{z}_{j} =-S+b^{\dagger}_{j}b_{j}$, $j=2,4$. Substituting these transformations into
 Hamiltonian (1), we get
 \begin{widetext}
 \begin{eqnarray}
 H&=&\sum_{\vec r}[
  (J_{\rm DF}S{\rm cos}2\theta-{1\over 2}J_{\rm D}{\rm sin}2\theta)
     (Y^{\dagger}_{1\vec r}+Y_{1\vec r}+Y^{\dagger}_{3\vec r}+Y_{3\vec r})
     +(J_{\rm DF}S{\rm sin}2\theta+J_{\rm D}{\rm cos}2\theta)
  (Y^{\dagger}_{1\vec r}Y_{1\vec r}+Y^{\dagger}_{3\vec r}Y_{3\vec r}) \nonumber\\
 &&+{1\over 2}J_{\rm D}(1+{\rm cos}2\theta)
  (u^{\dagger}_{1\vec r}u_{1\vec r}+d^{\dagger}_{1\vec r}d_{1\vec r}
    +u^{\dagger}_{3\vec r}u_{3\vec r}+d^{\dagger}_{3\vec r}d_{3\vec r})
    +(2J_{\rm a}S+{1\over 2}J_{\rm DF}{\rm sin}2\theta)
 (a^{\dagger}_{1\vec r}a_{1\vec r}+b^{\dagger}_{2\vec r}b_{2\vec r}
    +a^{\dagger}_{3\vec r}a_{3\vec r}+b^{\dagger}_{4\vec r}b_{4\vec r}) \nonumber\\
 &&-{\sqrt {2S}\over 2}J_{\rm DF}{\rm sin}(\theta+{\pi\over 4})
 (a^{\dagger}_{1\vec r}u^{\dag}_{1\vec r}+a^{\dagger}_{3\vec r}u^{\dag}_{3\vec r}
   +b_{2\vec r}d_{1\vec r}+b_{4\vec r}d_{3\vec r}+h.c.) \nonumber\\
 &&-{\sqrt {2S}\over 2}J_{\rm DF}{\rm sin}(\theta-{\pi\over 4})
 (a^{\dagger}_{1\vec r}d_{1\vec r}+a^{\dagger}_{3\vec r}d_{3\vec r}
   +u^{\dag}_{1\vec r}b_{2\vec r}+u^{\dag}_{3\vec r}b_{4\vec r}+h.c.)\nonumber\\
 &&+J_{\rm a}S(a_{3\vec r}b_{2\vec r}+h.c.)
 +{1\over 2}J_{\rm a}S(a_{1\vec r}b_{4\vec r-\vec b}+a^{\dag}_{1\vec r+\vec b}b^{\dag}_{4\vec r}+h.c.)
 +{1\over 2}J_{\rm a}S(a_{1\vec r}b_{2\vec r+\vec a}+a_{3\vec r}b_{4\vec r-\vec a}
  +a^{\dag}_{1\vec r-\vec a}b^{\dag}_{2\vec r}+a^{\dag}_{3\vec r+\vec a}b^{\dag}_{4\vec r} +h.c.)] \nonumber\\
 &&-2NJ_{\rm DF}S{\rm sin}2\theta-N({1\over 2}+{\rm
  cos}2\theta)J_{\rm D}-4NJ_{\rm a}S^2
  +{\rm3-operator\hspace{3pt}terms}+{\rm4-operator\hspace{3pt}terms}.
 \end{eqnarray}
 \end{widetext}
 The 3-operator and 4-operator terms are omitted here for simplicity. By letting the coefficient of
 the single-operator terms be zero, we get
 \begin{eqnarray}
 \theta=\frac{1}{2}{\rm arctan}\frac{2J_{\rm DF}S}{J_{\rm D}}.
 \end{eqnarray}
 Correspondingly, at zero temperature, we get the staggered magnetization at the Cu site as
 \begin{eqnarray}
 m_{\rm Cu}=|<T_{1z}>|={1\over 2}{\rm sin}2\theta=
 \frac{\frac{J_{\rm DF}}{J_{\rm D}}S}{\sqrt{1+(2\frac{J_{\rm DF}}{J_{\rm D}}S)^{2}}}.
 \end{eqnarray}
 When $J_{\rm DF}<<J_{\rm D}$, $\theta \approx \frac{J_{\rm DF}S}{J_{\rm D}}$
 and $m_{\rm Cu}\approx \frac{J_{\rm DF}S}{J_{\rm D}}$.
 This result is consistent with $m_{\rm Cu}\approx {J_{\rm DF}\over J_{\rm D}}m_{\rm Fe}$, which we get
 from the molecular field approximation in section II.

 After Fourier transformation, the Hamiltonian can be written as
 \begin{eqnarray}
 H&=&\sum_{k}
 \omega^{Y}_{k}(Y^{\dagger}_{1,k}Y_{1,k}+Y^{\dagger}_{3,k}Y_{3,k})
 +\Psi^{\dagger}_{k}H_k\Psi_{k}+C,\nonumber
 \end{eqnarray}
 where,
 \begin{eqnarray}
 \omega^Y_k&=&J_{\rm DF}S{\rm sin}2\theta+J_{\rm D}{\rm cos}2\theta,\nonumber\\
 C&=&-2NJ_{\rm DF}S{\rm sin}2\theta-N({1\over 2}+{\rm cos}2\theta)J_{\rm D}-4NJ_{\rm a}S^2,\nonumber\\
 \Psi^{\dagger}_k&=&(a^{\dagger}_{1,k},b_{2,-k},a^{\dagger}_{3,k},b_{4,-k},
 d^{\dagger}_{1,k},u_{1,-k},d^{\dagger}_{3,k},u_{3,-k}),\nonumber
 \end{eqnarray}

 \begin{widetext}
 \begin{eqnarray}
 H_k&=&\left(\begin{matrix}
 a & \delta e^{-ik_{y}} & 0 & \delta e^{ik_{x}} & -\gamma{\rm cos}\nu & \gamma{\rm sin}\nu & 0 & 0\\
 \delta e^{ik_{y}} & a & \delta & 0 & \gamma{\rm sin}\nu &-\gamma{\rm cos}\nu & 0 & 0\\
 0 & \delta & a & \delta e^{ik_{y}} & 0 & 0 &  -\gamma{\rm cos}\nu & \gamma{\rm sin}\nu\\
 \delta e^{-ik_{x}} & 0 & \delta e^{-ik_{y} } & a & 0 & 0 & \gamma{\rm sin}\nu &-\gamma{\rm cos}\nu\\
 -\gamma{\rm cos}\nu & \gamma{\rm sin}\nu & 0 & 0 & d & 0 & 0 & 0\\
 \gamma{\rm sin}\nu & -\gamma{\rm cos}\nu & 0 & 0 & 0 & d & 0 & 0\\
 0 & 0 & -\gamma{\rm cos}\nu & \gamma{\rm sin}\nu & 0 & 0 & d & 0\\
 0 & 0 & \gamma{\rm sin}\nu & -\gamma{\rm cos}\nu & 0 & 0 & 0 & d
 \end{matrix}\right)\label{ham}
 \end{eqnarray}
 \end{widetext}
 with
 \begin{eqnarray}
 a&=&2J_{\rm a}S+{1\over 2}J_{\rm DF}{\rm sin}2\theta,\nonumber\\
 d&=&{1\over 2}J_{\rm D}(1+{\rm cos}2\theta),\nonumber\\
 \gamma&=&-\sqrt{\frac{S}{2}}J_{\rm DF},\nonumber\\
 \delta&=&J_{\rm a}S,\nonumber\\
 \nu&=&\theta+\frac{\pi}{4}.
 \end{eqnarray}

 Compared with the conventional spin wave theory, the magnon gets an additional energy of
 ${1\over 2}J_{\rm DF}{\rm sin}2\theta=J_{\rm DF} m_{\rm Cu}$, which comes from the polarization of
 the Cu$^{2+}$ spin dimer. As a result, the magnon excitations
 show a tiny gap. We will discuss it later. The Hamiltonian can be diagonalized as
 \begin{eqnarray}
 H&=&\sum_{k}\sum_{i=1,8}\omega_{ik}\alpha_{ik}^{\dagger}\alpha_{ik}
    +\sum_k\omega^Y_{k}(Y^{\dagger}_{1k}Y_{1k}+Y^{\dagger}_{3k}Y_{3k}),\nonumber
 \end{eqnarray}
 where, a constant has been neglected and the expressions for $\omega_{ik}$ are given in Appendix A.

 The triplon excitations $Y_{\sigma k}$ ($\sigma=1,3$) do not interact with the magnons.
 They are dispersionless. The spectrum
 $\omega^Y_k=J_{\rm D}[1+2({J_{\rm DF}S\over J_{\rm D}})^2]/\sqrt{1+4({J_{\rm DF}S\over J_{\rm D}})^2}
 \approx J_{\rm D}[1+2({J_{\rm DF}S\over J_{\rm D}})^4]$. It increases with increasing $J_{\rm DF}$ and
 returns to the singlet-triplet gap when $J_{\rm DF}=0$. With the Cu$^{2+}$ spin operator (eq. (6))
 expressed by the $X_{\sigma}$ and $Y_{\sigma}$ (eq. (10)) and making mean-field approximation of
 $<X_{\sigma}>=1$, we find $T^z$ proportional to $Y$ and $Y^{\dag}$ and $T^{\pm}$ ($T^{x(y)}$)
 proportional to $u(u^{\dag})$ and $d(d^{\dag})$. Thus the operators
 $Y_{\sigma k}={\rm sin}\theta s_{\sigma k}+{\rm cos}\theta t_{\sigma k}$ ($\sigma=1,3$) describe
 the longitudinal fluctuations and we believe they are the modes
 {\it having longitudinal polarization and totally incompatible with conventional spin wave theory},
 as guessed by Masuda et. al\cite{masuda2004}.
 It is emphasized that these modes cannot be obtained by spin wave theory or
 effective spin wave theory\cite{masuda2007}.

 Among the eight $\alpha_{ik}$ excitations, four branches are magnon-like and another four are triplon-like. We denote
 them by $A_{1k}$, $A_{3k}$, $B_{2k}$, $B_{4k}$ and $D_{1k}$, $D_{3k}$, $U_{1k}$, $U_{3k}$ respectively. The
 corresponding spectra are represented by $\omega^{A_{1}}_{k}$, $\omega^{A_{3}}_{k}$, $\omega^{B_{2}}_{k}$,
 $\omega^{B_{4}}_{k}$ and $\omega^{D_{1}}_{k}$, $\omega^{D_{3}}_{k}$, $\omega^{U_{1}}_{k}$, $\omega^{U_{3}}_{k}$.
 They are two-fold degenerate. At $J_{\rm DF}=0$, they return to the pure magnon spectra
 $\omega_k^{A/B}=J_{\rm a}S\sqrt{2(1\mp {\rm cos}\frac{k_x}{2})}$ and pure triplon spectra $\omega_k^{U/D}=J_{\rm D}$.
 The magnons are gapless and the triplons are dispersionless. At a finite $J_{\rm DF}$,
 the magnons become gapful and the triplons become mobile due to the magnon-triplon interaction.

 Experimentally, $J_{\rm D}$ and $J_{\rm a}$ were determined as $J_{\rm D}=24$ meV and $J_{\rm a}=1.6$ meV.
 The reported $J_{\rm DF}$ is discrepant such as $2.4$ meV, $0.9$ meV  or $2.0$ meV. In Fig. 3 and 4,
 we show the numerically calculated magnon-like and triplon-like excitation spectra, respectively,
 with $J_{\rm D}=24$ meV, $J_{\rm a}=1.6$ meV
 and $J_{\rm DF}=1$ meV. The calculated magnon gap is $1.27$ meV. Experimentally, a small empirical anisotropy
 gap is estimated as $1$ meV  (spin wave theory\cite{masuda2007}) or
 $2.02$ meV  (effective spin wave theory\cite{masuda2004}). We find that the gap largely depends on $J_{\rm DF}$.
 The triplon-like excitations have a dispersion with small amplitude.

 \begin{figure}
 \begin{center}
 \includegraphics[ scale=0.70 ]{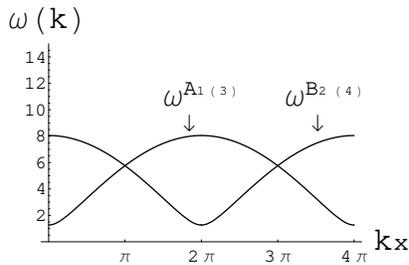}\\
 \caption{Spectra of the magnon-like excitations $\omega^{A_{1}}_{k}$, $\omega^{A_{3}}_{k}$, $\omega^{B_{2}}_{k}$,
 $\omega^{B_{4}}_{k}$ with $J_{\rm D}=24$ meV , $J_{\rm a}=1.6$ meV  and $J_{\rm DF}=1$ meV.
 Compared to the case of $J_{\rm DF}=0$, a small gap opens.}\label{sw2d}
 \end{center}
 \end{figure}

 \begin{figure}
 \begin{center}
 \includegraphics[ scale=0.70 ]{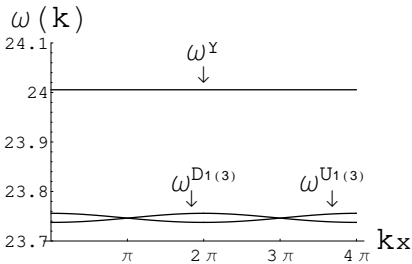}\\
 \caption{Spectra of the triplon-like excitations $\omega^Y_k$, $\omega^{D_{\sigma}}_{k}$ and
 $\omega^{U_{\sigma}}_{k}$ ($\sigma=1,3$) with $J_{\rm D}=24$ meV , $J_{\rm a}=1.6$ meV  and $J_{\rm DF}=1$ meV.
 The dispersionless $\omega^Y_k$ is longitudinally polarized, as mentioned by Masuda et. al.\cite{masuda2004}
 Compared to the case of $J_{\rm DF}=0$, $\omega^{D_{\sigma}}_k$ and $\omega^{U_{\sigma}}_k$ ($\sigma=1,3$)
 show weak dispersions.}\label{d2d}
 \end{center}
 \end{figure}

 Regarding $J_{\rm DF}\over J_{\rm D}$ as a small amount, we could get analytical expressions for the spectra. The
 details are given in the Appendix.
 The magnon-like excitation has a gap $\Delta_A=\omega^{A_{\sigma}}(0,\pi)=\Delta_B=\omega^{B_{\sigma^{\prime}}}(2\pi,\pi)
 \approx J_{D}\sqrt{\frac{2\eta}{1-\eta^2}}|\rho|$ ($\sigma=1,3$, $\sigma^{\prime}=2,4$)
 with $\eta=\frac{2J_{\rm a}S}{J_{\rm D}}$ and $\rho=-\frac{J_{\rm DF}}{J_{\rm D}}\sqrt{\frac{S}{2}}$.
 It should be pointed out that the magnon-like spectra are proportional to $({J_{\rm DF}\over J_{\rm D}})^2$
 away from the momentum of $(0,\pi)$ or $(2\pi,\pi)$ (see eq. (\ref{ab})).
 The gap of the triplon-like excitations is
 $\Delta_U=\omega^{U_{\sigma}}(2\pi,\pi)=\Delta_D
 =\omega^{D_{\sigma}}(0,\pi)\approx J_{\rm D}(1-\frac{\eta}{1-\eta^2}\rho^2)$
 ($\sigma=1,3$). When $\eta<<1$, the gaps can be further simplified as
 $\Delta_A\approx J_{\rm DF}S\sqrt{\frac{2J_{\rm a}}{J_{\rm D}}}\approx J_{\rm DF}$
 and $\Delta_U\approx J_{\rm D}[1-{{J_{\rm a}\over J_{\rm D}}({{J_{\rm DF}S} \over J_{\rm D}})^2}]
 \propto ({J_{\rm DF}\over J_{\rm D}})^2$.
 These relations are numerically shown in Fig. 5.
 Reminding $\Delta_Y\propto ({J_{\rm DF}S\over J_{\rm D}})^4$, we find that different branches have different
 dependences on $J_{\rm DF}\over J_{\rm D}$. The band width of the triplon-like excitations is
 $W_U=W_D\approx J_{\rm D}\frac{\eta}{1-\eta^2}\rho^2\approx J_{\rm a}({J_{\rm DF}S\over J_{\rm D}})^2$. As a counterpart,
 an effective interaction between the Cu$^{2+}$ dimers, mediated by the Fe$^{3+}$ chains,
 can be estimated as $J_{\rm eff}^{\prime}\sim J_{\rm a}({J_{\rm DF}S\over J_{\rm D}})^2$.

Experimentally\cite{masuda2004,masuda2007}, the magnetic moment in Cu$_{2}$Fe$_{2}$Ge$_{4}$O$_{13}$
 was found to be nearly confined in the a-c plane. Dzyaloshinskii-Moriya (DM) interaction and other
 insignificant anisotropy effects may exist. For simplicity, we neglect these anisotropy effects and
 the weak interaction along c-direction in our mean-field theory.
 Some remarks on the Goldstone theorem have to be made here. We start from an isotropic spin Hamiltonian and
 a gapless mode is expected in the ordered state because of the rotational symmetry. The small but finite magnon
 gap seems unwelcome. We interpret it by a two-step process. Similar to the mixed-spin antiferromagnets
 R$_2$BaNiO$_5$\cite{zheludev01}, there is a separation of energy scales of magnetic excitations in
 Cu$_{2}$Fe$_{2}$Ge$_{4}$O$_{13}$. The Fe$^{3+}$ centered magnons with low frequencies and the Cu$^{2+}$ dimer
 centered triplons with high frequencies have different dynamical behaviors and different timescales.
 At the mean-field level, we first consider the effective Fe-Fe interaction $J_{\rm eff}$ and the induced long range order
 in the Fe$^{3+}$ sublattices. Goldstone modes appear at this stage. We then study the polarization of the
 Cu$^{2+}$ dimers and its reaction on the Fe$^{3+}$ sublattices. The hybridized magnons get a small gap due to
 the staggered field presented by the polarized Cu$^{2+}$ dimers.
 This phenomenon has been studied earlier in spin-${1\over 2}$ and spin-$1$ antiferromagnetic
 spin chains experimentally\cite{zheludev96,zheludev98,dender}
 and theoretically\cite{affleck}. It is shown that the presence of a staggered field will
 make the spin-${1\over 2}$ antiferromagnetic chain gapful and split the Haldane triplet into two branches.
 In our studies, we further consider the fluctuations of the staggered field and their hybridization
 with the magnons. It may be interesting to consider the DM interaction or other anisotropic effects,
 which break the rotational symmetry and lead a gap naturally. We argue that the polarized Cu$^{2+}$
 dimers will make important contributions to the gap, even dominant if the anisotropies are tiny.
 We note that the estimated gap is close to the estimated $J_{\rm DF}$ from either
 the spin wave theory\cite{masuda2007} or the effective spin wave theory\cite{masuda2004}.

In the case of $J_{\rm DF}<0$, the Cu$^{2+}$ spin $\hat T_n$ will be parallel to its nearest
 Fe$^{3+}$ spin $\hat S_n$ ($n=1,2,3,4$). Our theory is still applicable. The angle $\theta$
 becomes $\theta^{\prime}=-{\frac{1}{2}} {\rm arctan} {\frac{2|J_{\rm DF}|S}{J_{\rm D}}}$.
 In Fig. 5, we also show the dependence of the
 energy gaps on ${J_{\rm DF}\over J_{\rm D}}$ in the range of $J_{\rm DF}<0$.

 \begin{figure}
 \begin{center}
 \includegraphics[ scale=0.60 ]{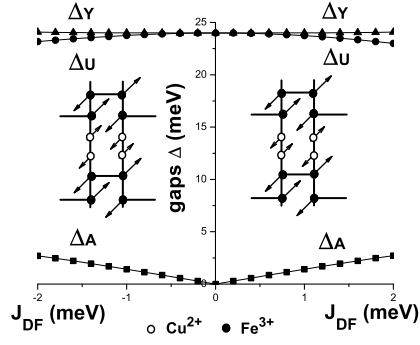}\\
 \caption{The magnon-like excitations have a gap $\Delta_A$ proportional to $J_{\rm DF}$(squares),
  while the triplon-like excitation shows $J_{\rm DF}^4$ ($\Delta_Y$, triangles) or
  $J_{\rm DF}^2$ dependences ($\Delta_U$, circles). $J_{\rm D}=24$ meV and $J_{\rm a}=1.6$ meV.
  When $J_{\rm DF}<0$, the Cu$^{2+}$ spin is parallel to its nearest Fe$^{3+}$ spin; while when
  $J_{\rm DF}>0$, they are antiparallel.}\label{pert02}
 \end{center}
 \end{figure}

\section{Summary}

 In summary, we combine the bond-operator representation and Holstein-Primakoff transformation
 to study the mixed spin lattice model made up of Cu$^{2+}$ ($S=\frac{1}{2}$)
 spin-dimers and Fe$^{3+}$ ($S=\frac{5}{2}$) spin chains. A finite interaction $J_{\rm DF}$ between the spin-dimer
 and the spin chain introduces a cooperative order. The effective interaction between the spin chains,
 mediated by the spin dimer, is calculated up to the third order. It makes the Fe$^{3+}$ spins ordered
 antiferromagnetically. Meanwhile, the local magnetization of Fe$^{3+}$ spins polarizes the Cu$^{2+}$ spin dimers
 and the staggered Cu$^{2+}$ magnetization is shown proportional to $J_{\rm DF}$ as well as the local Fe$^{3+}$
 magnetization. It effectively presents a staggered field reacting on the Fe$^{3+}$ spin chains.
 Considering the triplon-magnon interaction, the excitation spectra are especially investigated.
 The degeneracy of the triplons is lifted and the mode {\it having longitudinal polarization and
 totally incompatible with conventional spin wave theory}, as pointed out by Masuda et. al, is identified.
 It shows $({J_{\rm DF}S\over J_{\rm D}})^4$ dependence near vanishing $J_{\rm DF}$.
 The hybridized triplon-like excitations get a weak dispersion and show a different
 $({J_{\rm DF}\over J_{\rm D}})^2$ dependence.
 At the same time, the hybridized magnon-like excitations
 open a gap $\Delta_A=\Delta_B\approx J_{\rm DF}S\sqrt{\frac{2J_{\rm a}}{J_{\rm D}}}\approx J_{\rm DF}$ at the momentum of
 $(0,\pi)$ or $(2\pi,\pi)$. Away from this momentum, the magnon-like spectra
 show $({J_{\rm DF}\over J_{\rm D}})^2$ relation, instead.
 The experiments on Cu$_{2}$Fe$_{2}$Ge$_{4}$O$_{13}$ are interpreted.

\section{Acknowledgements} We acknowledge the financial support by
 Natural Science Foundation of China and $973$-project under
 grant no. 2006CB921300 and 2006CB921400.

\begin{appendix}

\section{excitation spectra}

From the matrix $H_k$ (eq. (\ref{ham})), we get the equation to
determine the excitation spectra:
 \begin{eqnarray}
 \omega^{8}+U\omega^{6}+V\omega^{4}+W\omega^{2}+Z=0,
 \end{eqnarray}
 where,
 \begin{widetext}
 \begin{eqnarray}
 U&=&-2( a^{2} + d^{2} - 2\delta^{2}+
    2\gamma^2{\rm cos}\varphi), \nonumber\\
 V&=&a^4 + 4a^2d^2 + d^4 - 4ad\gamma^2 + 3\gamma^4 - 4a^2\delta^2 -
  8d^2\delta^2 + 2\delta^4 -
  2\delta^4{\rm cos}k_{x}+4\gamma^2( a^2 + d^2 - 2\delta^2)
     {\rm cos} \varphi \nonumber\\
  &&+3{\gamma }^4{\rm cos}^2\varphi -
  4d\gamma^2\delta{\rm cos}k_{y}
   {\rm sin}\varphi - 3\gamma^4{\rm \sin}^{2}\varphi,\nonumber\\
 W&=&-2a^4d^2 + 4a^3d\gamma^2 +
  4d^4\delta^2 + 2\gamma^4\delta^2 +
  4ad\gamma^2( d^2 - 2\delta^2 )-a^2( 2d^4 + \gamma^4-8d^2\delta^2 ) -
  d^2( \gamma^4 + 4\delta^4 )\nonumber\\
  &&+4d\delta^3{\rm cos}k_{x}( d\delta+\gamma^2{\rm cos}k_{y}{\rm sin}\varphi)
  + \gamma^2( -( \gamma^2( a^2 + d^2 - 2\delta^2 ) {\rm cos} (2\varphi ))  -
     \gamma^4{\rm cos}(3\varphi )\nonumber\\
  && +4d\delta( a^2 + d^2 -\delta ^2 ){\rm cos}k_{y}{\rm sin}\varphi
     +{\rm cos}\varphi( -4a^2d^2 +8ad\gamma^2 - 3\gamma^4 +
        8d^2\delta^2 +8d\gamma^2\delta{\rm cos}k_{y}{\rm sin}\varphi)) \nonumber\\
 Z&=&a^4d^4 - 4a^3d^3\gamma^2 +\frac{3\gamma^8}{8} -
  \frac{5d^2\gamma^4\delta^2}{2} +2d^4\delta^4+a^2( 5d^2\gamma^4-4d^4\delta^2)
  +2a( -d\gamma^6+ 4d^3\gamma^2\delta^2)\nonumber\\
  &&-d^2\delta^2{\rm cos}k_{x}( \gamma^4 + 2d^2\delta^2 -\gamma^4{\rm cos}(2\varphi ))
  +\frac{1}{8}\gamma^4( 4( 2a^2d^2 - 4ad\gamma^2 +\gamma^4
  - 3d^2\delta^2){\rm cos}(2\varphi )+\gamma^4{\rm cos} (4\varphi ) ) \nonumber\\
  &&-2d\gamma^2\delta{\rm cos}k_{y}( 2a^2d^2 - 4ad\gamma^2 +
     \gamma^4 - 2d^2\delta^2 + 2d^2\delta^2{\rm cos}k_{x}
     +\gamma^4{\rm cos}(2\varphi )){\rm sin}\varphi
    + d^2\gamma^4\delta^2( 1 + 2{\rm cos}(2k_{y})){\rm sin}^2
    \varphi,\nonumber
 \end{eqnarray}
 with $\varphi=2(\theta+\frac{\pi}{4})$.

 Its roots are
 \begin{eqnarray}
 \omega^{2}_{\epsilon,\sigma}=-\frac{1}{4}(U+\epsilon\sqrt{8\xi+U^{2}-4V})
    +\sigma\sqrt{\frac{1}{16}(U+\epsilon\sqrt{8\xi+U^{2}-4V})^{2}
    -(\xi+\epsilon\frac{U\xi-W}{\sqrt{8\xi+U^{2}-4V}})},
 \end{eqnarray}
 where $\epsilon=\pm 1$, $\sigma=\pm 1$ and
 \begin{eqnarray}
 \xi=\frac{V}{6}+\left(-\frac{q}{2}+\sqrt{\left(\frac{q}{2}\right)^{2}
    +\left(\frac{p}{3}\right)^{3}}\right)^{1/3}
    +\left(-\frac{q}{2}-\sqrt{\left(\frac{q}{2}\right)^{2}
    +\left(\frac{p}{3}\right)^{3}}\right)^{1/3},
 \end{eqnarray}
 with $p=\frac{UW}{4}-\frac{V^{2}}{12}-Z$ and
 $q=\frac{UVW}{24}-\frac{V^{3}}{108}-\frac{W^{2}}{8}-\frac{U^{2}Z}{8}+\frac{VZ}{3}$.
 If $|J_{\rm DF}|\ll J_{\rm D}$, the triplon-like excitation spectra can be expanded as
 \begin{eqnarray}
 \omega^{U_{\sigma}}_{k}&=&J_{\rm D}{\Big(}1+g_{1}(k)\rho^2{\Big)},\nonumber\\
 \omega^{D_{\sigma}}_{k}&=&J_{\rm D}{\Big(}1+g_{-1}(k)\rho^2{\Big)},~~~~\sigma=1,3,
 \end{eqnarray}
 where, $\rho=-\frac{J_{\rm DF}}{J_{\rm D}}\sqrt{\frac{S}{2}}$ and
 \begin{eqnarray}
 g_{\epsilon}(k)=\frac{\eta{\Big(}-4\eta^2+8-{\rm cos}k_{x}(3\eta^2-4+\eta^2{\rm cos} k_{y})
 +2\epsilon(\eta^2+(\eta^2-1){\rm cos}k_{x})\sqrt{3+{\rm
 cos}(2k_{y})}{\Big)}}{-\eta^4+8\eta^2-8+\eta^4{\cos}k_{y}}
 \end{eqnarray}
 \end{widetext}
 with $\eta=\frac{2J_{\rm a}S}{J_{\rm D}}$. The energy gap of the triplon-like excitations
 \begin{eqnarray}
 \Delta_U&=&\Delta_D \approx
 J_{\rm D}(1-\frac{\eta}{1-\eta^2}\rho^2),
 \end{eqnarray}
 locating at the momentum of $(2\pi,\pi)$ (for the branch of
 $\omega^{U_{\sigma}}$, $\sigma=1,3$) or
 $(0,\pi)$ (for the branch of $\omega^{D_{\sigma}}$, $\sigma=1,3$).
 The band width of the triplon-like excitations is
 \begin{eqnarray}
 W_U=W_D \approx J_{\rm D}\frac{\eta}{1-\eta^2}\rho^2.
 \end{eqnarray}

 The energy gap of the magnon-like excitations occurs at the momentum of
 $(0,\pi)$ (for the branch of $\omega^{A_\sigma}$, $\sigma=1,3$)
 or $(2\pi,\pi)$ (for the branch of $\omega^{B_\sigma}$, $\sigma=2,4$)
 and shows linear dependence on $J_{\rm DF}$:
 \begin{eqnarray}
 \Delta_A=\Delta_B=J_{\rm D}\sqrt{\frac{2\eta}{1-\eta^2}}|\rho|
 \approx |J_{\rm DF}|S\sqrt{\frac{2J_{\rm a}}{J_{\rm D}}}|\approx
 |J_{\rm DF}|.\nonumber\\
 \end{eqnarray}
 Away from that momentum, the magnon-like excitation spectra can be expanded as
 \begin{eqnarray}
 \omega^{A_{\sigma}}_{k}&=&J_{\rm D}{[}\sqrt{\frac{1}{2}\eta^2(1-{\rm cos}\frac{k_x}{2})
 }+g_{1}(k)\rho^2{]},\nonumber\\
 \omega^{B_{\sigma+1}}_{k}&=&J_{\rm D}{[}\sqrt{\frac{1}{2}\eta^2(1+{\rm cos}\frac{k_x}{2})
 }+g_{-1}(k)\rho^2{]},~~~~\sigma=1,3.\nonumber\\\label{ab}
 \end{eqnarray}
 Interestingly, they exhibit quadratic dependence on $J_{\rm DF}\over J_{\rm D}$.
 The band width of the magnon-like excitations is
 \begin{eqnarray}
 W_A=W_B \approx J_{\rm D}{\Big(}\eta-\sqrt{\frac{2\eta}{1-\eta^2}}|\rho|{\Big)}.
 \end{eqnarray}

\end{appendix}

\end{document}